\documentclass{article}
\usepackage{arxiv}

\usepackage[utf8]{inputenc} 
\usepackage[T1]{fontenc}    
\usepackage{hyperref}       
\usepackage{url}            
\usepackage{booktabs}       
\usepackage{amsfonts}       
\usepackage{nicefrac}       
\usepackage{microtype}      
\usepackage{lipsum}
\usepackage{graphicx}

\graphicspath{ {./images/} }
\mathchardef\mhyphen="2D
\usepackage[super,sort&compress,comma]{natbib} 
\usepackage{fancyhdr}

\title{Hybridizing physical and data-driven prediction \mbox{methods} for physicochemical properties}

\author{
 Fabian Jirasek\thanks{Present address: Laboratory of Engineering Thermodynamics (LTD), TU Kaiserslautern, Germany. E-mail: fabian.jirasek@mv.uni-kl.de.} \\
  Department of Computer Science\\
  University of California\\
  Irvine, CA 92697, USA\\
   \And
 Robert Bamler \\
  Department of Computer Science\\
  University of California\\
  Irvine, CA 92697, USA\\
  \And
 Stephan Mandt \\
  Department of Computer Science\\
  University of California\\
  Irvine, CA 92697, USA\\
}

\begin{document}
\maketitle
\begin{abstract}
We present a generic way to hybridize physical and data-driven methods for predicting physicochemical properties. The approach `distills' the physical method's predictions into a prior model and combines it with sparse experimental data using Bayesian inference. We apply the new approach to predict activity coefficients at infinite dilution and obtain significant improvements compared to the data-driven and physical baselines and established ensemble methods from the machine learning literature.  
\end{abstract}

Prediction methods for physicochemical properties are indispensable for process design and optimization in chemical engineering since experimental studies are expensive and tedious. The most widely used approaches are group-contribution methods (GCMs) that model the properties of pure components or mixtures based on the structural groups that build up the components.~\cite{fredenslund1989,constantinou1994,nannoolal2004,gardas2009} GCMs can also be used for predicting properties of mixtures of which the composition is unknown.~\cite{jirasek2018,jirasek2019,jirasek2020a} The most successful GCMs for mixtures are the different versions of UNIFAC~\cite{fredenslund_1975,fredenslund_1977,weidlich_1987} that model the excess Gibbs energy based on binary group-interaction parameters. The group-contribution concept greatly reduces the number of model parameters and the amount of data needed for fitting GCMs. However, the practical applicability of UNIFAC is still restricted, mainly due to necessary group-interaction parameters that have not been fitted yet.

Another successful approach is the quantum chemistry-based COSMO-RS~\cite{klamt_1995}, which describes the properties of mixtures referring to the polarization charge densities of the constituent components, and which depends only on a small number of adjustable parameters.~\cite{klamt2010} However, expensive COSMO calculations are required for each component. \\ 
In previous work~\cite{jirasek_2020}, we have introduced a novel, purely data-driven approach to predict physicochemical properties of mixtures. Specifically, we considered activity coefficients at infinite dilution $\gamma_{ij}^\infty$ in binary mixtures at a constant temperature, but this approach generalizes to other properties. The data for $\gamma_{ij}^\infty$ can be represented as a matrix whose rows and columns correspond to solutes~$i$ and solvents~$j$, respectively.
For $\gamma_{ij}^\infty$ at $298.15\pm1$~K, which we studied in our previous work, the matrix containing the available experimental data from one of the largest databases for physicochemical properties, the Dortmund Data Bank~\cite{DDB_2019}, is very sparse, cf. Figure~\ref{Fig:S1} in the Supporting Information. The data set covers 240 solutes and 250 solvents, but only 4{,}094 entries are observed. The prediction of the unobserved entries, i.e., the prediction of $\gamma_{ij}^\infty$ for not yet studied mixtures, can be framed as a matrix completion problem~\cite{candes2009,mazumder2010}.

The basis of our previously introduced approach~\cite{jirasek_2020} is a probabilistic matrix completion method (MCM). We modeled  
$\ln\gamma_{ij}^\infty$ (the logarithm of $\gamma_{ij}^\infty$ is used for scaling purposes) as a stochastic function of initially unknown features of the solutes $i$ and solvents $j$, specifically as the dot product of two vectors:
\begin{equation}
    \ln\gamma_{ij}^\infty = u_i \cdot v_j + \epsilon_{ij}
    \label{Eq.MCM}
\end{equation}
where $u_i$ and $v_j$ are learned feature vectors for solute~$i$ and solvent~$j$, respectively, and the random variable $\epsilon_{ij}$ captures both measurement noise and inaccuracies of the model. The feature vectors of all considered solutes and solvents can be aggregated to two feature matrices $U$ and~$V$, respectively. Rather than selecting features based on physical considerations, the data-driven approach infers useful features from available experimental data on $\ln\gamma_{ij}^\infty$ alone, using the laws of probability theory and (approximate) Bayesian inference.~\cite{murphy2012machine, blei2017variational, kucukelbir2017} The inferred features can then be used to predict $\ln\gamma_{ij}^\infty$ for mixtures for which no experimental data are available, cf. Eq.~(\ref{Eq.MCM}).

While the purely data-driven approach~\cite{jirasek_2020} already outperforms the state-of-the-art physical method for predicting activity coefficients modified UNIFAC (Dortmund)~\cite{weidlich_1987,constantinescu_2016} (to which we simply refer as UNIFAC in the following) in terms of average predictive performance, it leaves substantial room for improvement as it ignores available physical knowledge about the mixtures. In thermodynamics, such knowledge is often abundant, e.g., in pure component properties or physical laws and models. In this paper, we therefore propose a hybrid physics-based/data-driven prediction method that combines the best of both worlds. We show that the framework of probabilistic models and Bayesian inference provides a principled way to incorporate scientific domain knowledge into machine learning (ML) models by specifying a so-called prior probability distribution over model parameters. Specifically, we propose to use \emph{model distillation} \cite{hinton2015distilling} to extract physical domain knowledge from UNIFAC in a format that can be used to construct an informative prior distribution for the MCM.\\
In the following, we describe the details of our proposed hybrid approach. Once again, we consider predicting $\ln\gamma_{ij}^\infty$ in binary mixtures at $298.15\pm 1$~K as a prime example and evaluate the predictive performance on the same data set as in our previous work~\cite{jirasek_2020}. As physical base method, we use the current publicly available version of UNIFAC~\cite{weidlich_1987,constantinescu_2016}. As data-driven base method, we adopt the Bayesian MCM from our previous work~\cite{jirasek_2020}. We compare the performance of our hybrid method to the performances of the constituent base methods as well as two established ML ensemble methods.

\begin{figure}[tb]
    \centering
    \includegraphics[width=0.7\textwidth]{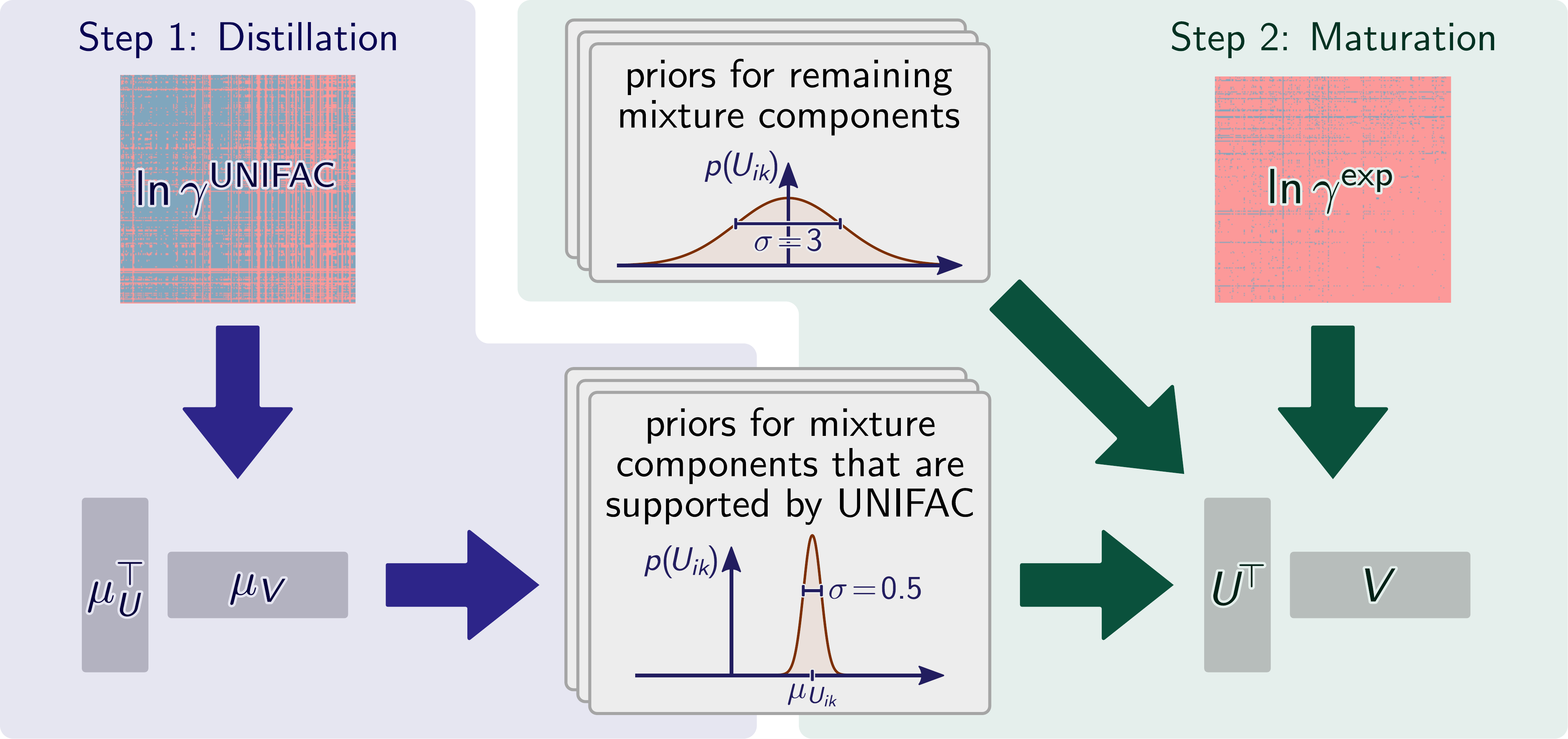}
    \caption{Scheme of the \emph{whisky} method. We first fit an MCM to UNIFAC predictions for $\ln\gamma_{ij}^{\infty}$ (\emph{distillation}, purple). We then use the fitted parameters from the distillation step to construct informative priors for the component feature matrices $U$ and~$V$ and fit the model to experimental data on $\ln\gamma_{ij}^{\infty}$ using these priors (\emph{maturation}, green). $\ln\gamma^{\rm UNIFAC}$ and $\ln\gamma^{\rm exp}$ denote the available data sets from UNIFAC~\cite{constantinescu_2016} and experiments~\cite{DDB_2019}, respectively.}
    \label{flowchart}
\end{figure}
Figure~\ref{flowchart} summarizes our proposed hybrid method, which we call \emph{whisky}.
Just like the manufacturing of whisky, our whisky method involves a \emph{distillation} step, in which we \emph{distill} knowledge from an existing model into a prior distribution using an approach known as \emph{model distillation} in the ML literature~\cite{hinton2015distilling}, and a \emph{maturation} step, in which we allow the prior to \emph{mature} by combining it with experimental data. Both steps are based on a probabilistic MCM similar to our previous work~\cite{jirasek_2020} to fit model parameters (i.e., feature matrices $U$ and~$V$) to a data set of $\ln\gamma_{ij}^{\infty}$. The difference between the distillation and maturation step is that they operate on different data sets. The distillation step fits an MCM to all predictions for $\ln\gamma_{ij}^{\infty}$ at 298.15~K that can be obtained with UNIFAC, denoted as $\ln\gamma^{\rm UNIFAC}$. Thus, the distillation step extracts the physical knowledge encoded in UNIFAC, which is implicitly exposed via its predictions for $\ln\gamma_{ij}^{\infty}$, into parameters of an MCM. By contrast, the maturation step builds upon the results of the distillation step and refines the parameters by fitting an MCM to the available experimental data, denoted as $\ln\gamma^{\rm exp}$.

The two different data sets $\ln\gamma^{\rm UNIFAC}$ and $\ln\gamma^{\rm exp}$ are illustrated in the two blue/red matrices in Figure~\ref{flowchart}. Here, rows and columns correspond to solutes and solvents, respectively, and blue or red entries indicate binary mixtures for which data points are available or absent, respectively. As can be seen, UNIFAC predictions are available for a lot more mixtures than experimental observations ($\ln\gamma^{\rm UNIFAC}$ has more blue entries than $\ln\gamma^{\rm exp}$, cf. Fig~\ref{Fig:S1} and \ref{Fig:S3} in the Supporting Information), meaning that the distillation step trains on a larger data set. While the experimental data set is more sparse, it is considered more reliable than the UNIFAC predictions.

The main novelty of our proposed whisky method lies in the way how it combines physical knowledge with experimental data. We realize the interface between distillation and maturation (purple and green parts of Figure~\ref{flowchart}) by specifying an informative prior distribution over model parameters. To understand the role of the prior, it is instructive to recall the principles of Bayesian inference on which our MCM builds. Bayesian inference describes the relationship between three probability distributions, called \emph{prior}, \emph{likelihood}, and \emph{posterior}. The prior is a probability distribution over model parameters that encodes a-priori knowledge, i.e., information on the model parameters before the model is fitted to the training data. In a purely data-driven approach, no a-priori information is used, and the prior is typically a very broad (i.e., noninformative) probability distribution. The likelihood encodes how model parameters manifest themselves in physically observable quantities, i.e., the data to which the model is trained. Together, prior and likelihood define a probabilistic model over observable quantities, such as $\ln\gamma_{ij}^\infty$ here. Bayesian inference takes such a probabilistic model and compares its predictions to actual observed data. The task of Bayesian inference is to find the so-called posterior probability distribution over model parameters that are consistent with the observed quantities \textit{and} the a-priori knowledge.

This framework of probabilistic modeling and Bayesian inference provides a principled way of hybridizing different methods using probability distributions as interfaces. Our approach, illustrated in Figure~\ref{flowchart}, follows the principle that `one man's ceiling is another man's floor'.
In analogy to this proverb, the posterior of the distillation step, which encodes knowledge after seeing the UNIFAC predictions, can be turned into a prior of the maturation step, which encodes knowledge before seeing the experimental data. Specifically, we construct a physically informed prior for the maturation step by taking the posterior means $\mu_U$ and $\mu_V$ from the distillation step, and we form Gaussian prior distributions with a rather small standard deviation of $\sigma=0.5$ around these means. Thus, this choice of prior encodes physical knowledge from the UNIFAC model. At the same time, the nonzero prior standard deviation allows the maturation step to overrule prior knowledge if the experimental data provide enough evidence to justify this.

For some of the considered mixture components (eight solutes and 41 solvents), UNIFAC is not applicable. Since the distillation step does not provide any information about these components, we use a broader (i.e., less informative) Gaussian prior here in the maturation step, with a standard deviation of $\sigma=3$ centered around zero. For the task of (approximate) Bayesian inference, we use the Stan framework~\cite{carpenter2017} and resort to variational inference~\cite{blei2017variational,kucukelbir2017}. More details on the proposed whisky method are given in the Supporting Information.

\begin{figure}[!htb]
    \centering
    \includegraphics[width=0.48\textwidth]{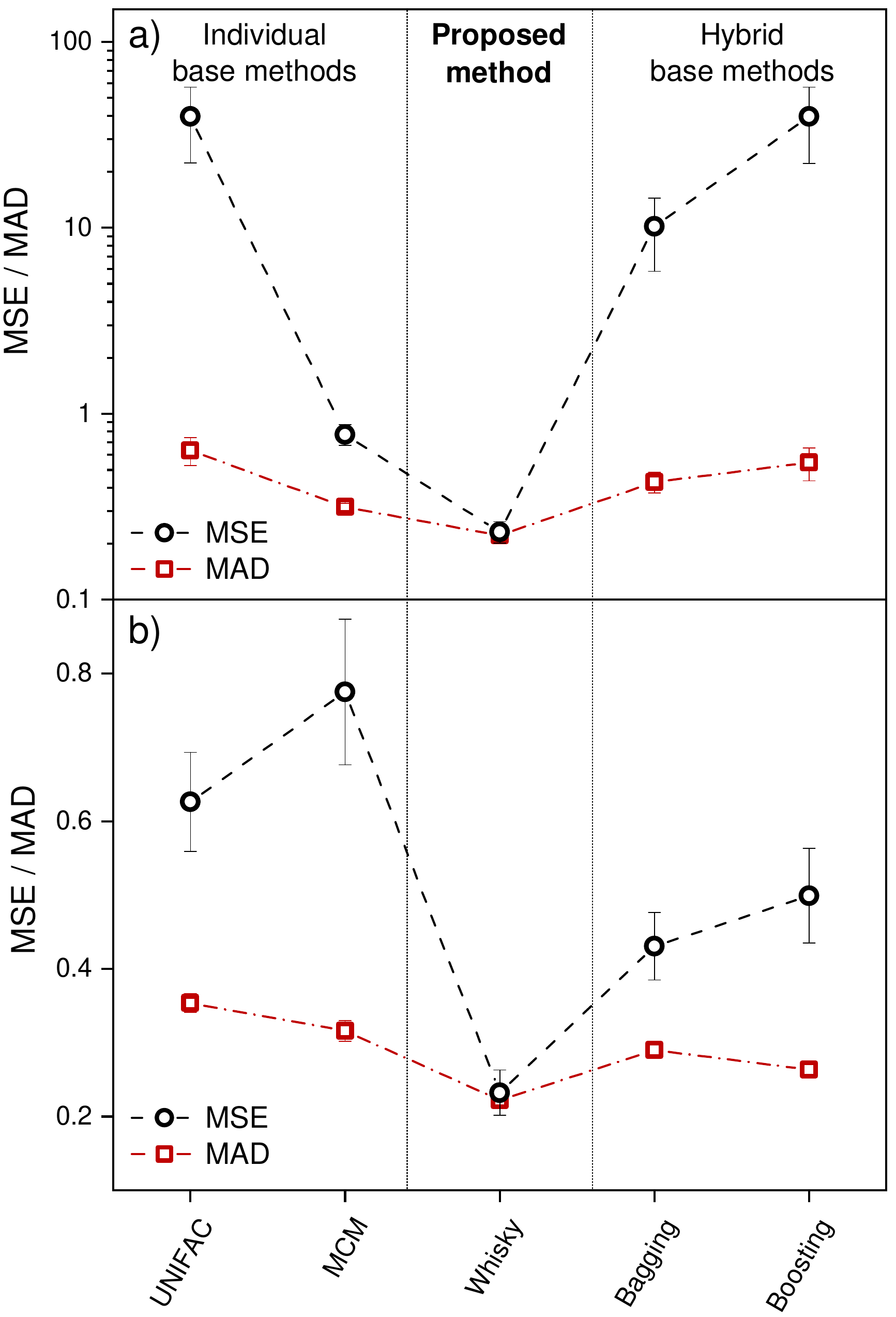}
\caption{Mean square error (MSE) and mean absolute deviation (MAD) for the prediction of $\ln\gamma_{ij}^\infty$ using the individual base methods UNIFAC and data-driven MCM, the proposed hybrid whisky method, and the hybrid baselines bagging and boosting. Lower is better for both metrics. Error bars show the standard errors of the means. a) Considering all applicable data points. b) Ignoring the worst eight outliers of UNIFAC.}
\label{Comparison}
\end{figure}

In Figure~\ref{Comparison}~a), we compare the overall performance of the whisky method for predicting $\ln\gamma_{ij}^\infty$ with the performances of the base methods UNIFAC~\cite{constantinescu_2016} and MCM~\cite{jirasek_2020} (without the informative prior), and with two alternative hybrid approaches, bootstrap aggregation (aka bagging)~\cite{breiman_1996} and boosting~\cite{schapire_1990}. We compare mean absolute deviation (MAD) and mean square error (MSE).

 Bagging is realized here by averaging the predictions from UNIFAC and the data-driven MCM for each data point; boosting is implemented by training an MCM to the matrix of the residuals of UNIFAC. Bagging and boosting are described in detail in the Supporting Information. To simulate predictive performances, the predictions with MCM, whisky, and boosting (and the \emph{MCM contribution} of bagging) are obtained by using leave-one-out cross-validation~\cite{hastie_2001}, i.e., by training the models to all experimental data points except for one, which is then used as a test data point and predicted. The training set of UNIFAC is not disclosed; hence, no statements on whether the UNIFAC results are obtained by regression or prediction can be made here.
 
Figure~\ref{Comparison}~a) demonstrates that the proposed whisky method outperforms all other methods in both MAD and MSE. The poor scores of UNIFAC, bagging, and boosting can mainly be attributed to only a handful of data points that are extremely poorly predicted by UNIFAC as shown in Figure~\ref{Fig:S8} in the Supporting Information. However, even if we, as an example, ignore the worst eight outliers of UNIFAC (marked in Figure~S.8) for the evaluation, the proposed whisky method still performs significantly better than all baselines, cf. Figure~\ref{Comparison}~b). \\
If the worst eight UNIFAC outliers are ignored (Figure~\ref{Comparison} b), the results show that the hybrid baselines -- bagging and boosting -- also improve the predictions of the base methods UNIFAC and MCM: bagging and boosting have smaller MAD and MSE values than the base methods. Bagging is widely used if the available base methods for a specific problem tend to overfit, i.e., if they fit the training data but do not generalize well to unobserved data.~\cite{hastie_2001} By contrast, boosting is commonly applied in ML to tackle the opposite problem of underfitting, which arises if the base methods are not expressible enough for a specific problem.~\cite{schapire_1990} The observation that our proposed whisky method performs better than both bagging and boosting indicates that the base methods UNIFAC and data-driven MCM tend to overfit to parts of the data set. At the same time, they also seem to underfit on other combinations, so that neither bagging nor boosting is universally applicable. This may in part be explained by the fact that the experimental data set is very imbalanced: while we have data for at least 86 different binary mixtures for each of the $5\%$ most common solutes, we only have six or fewer data points for each of the 50\% most uncommon solutes (see also Figure~\ref{Fig:S1} in the Supporting Information). The proposed whisky approach seems more robust to such an imbalanced data set than the other hybrid approaches.

\begin{figure}[!htb]
    \centering
    \includegraphics[width=0.48\textwidth]{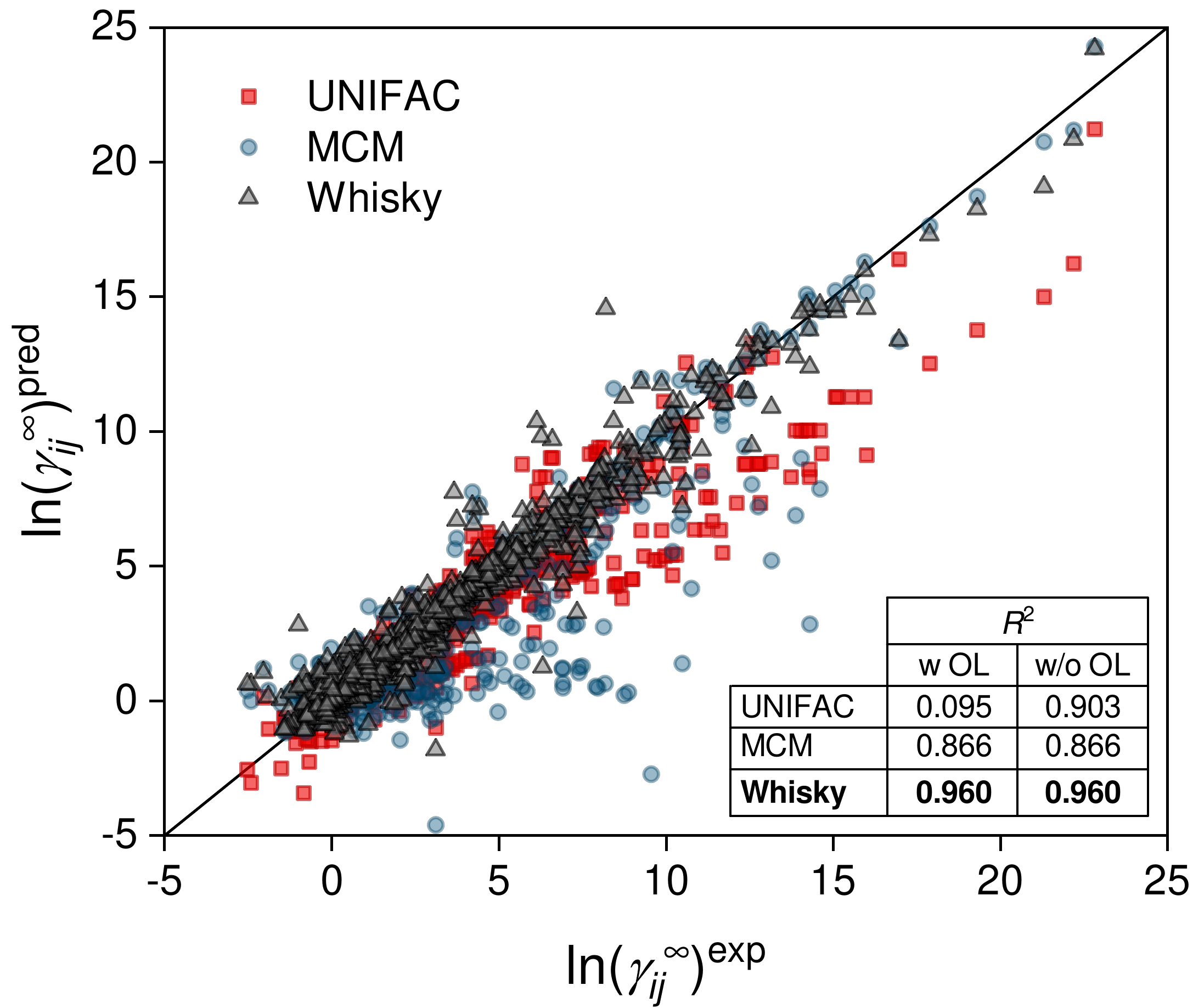}
    \caption{Parity plot of the predictions (pred) for $\ln\gamma_{ij}^\infty$ with the proposed whisky method over the corresponding experimental values (exp) and comparison to UNIFAC and data-driven MCM. Coefficients of determination~$R^2$ (higher is better, $1$ implies perfect correlation) are given, both including and excluding the worst eight UNIFAC outliers (OL).}
    \label{Parity_Plot}
\end{figure}

In Figure~\ref{Parity_Plot}, we compare the predictions of the whisky method with those of the data-driven MCM and UNIFAC in a parity plot. Points on the diagonal line correspond to perfect predictions. The whisky method reliably reduces outliers of both base methods. By contrast, both bagging and boosting, shown in Figure~\ref{Fig:S7}in the Supporting Information, only partially compensate for outliers of the data-driven MCM but severely suffer from outliers of UNIFAC. The whisky method also yields the highest coefficient of determination $R^2$ (with $R^2=1$ being optimal) of all compared methods, irrespective of whether the worst eight UNIFAC outliers~(OL) are considered or not (see table insets).\\
Another major advantage of the proposed whisky method is its broader applicability compared to the other hybrid approaches. For a fair comparison, Figures~\ref{Comparison} and \ref{Parity_Plot} consider only data points that can be predicted with UNIFAC, which is also a prerequisite for applying bagging and boosting. By contrast, the whisky method (and the data-driven MCM) can be used to predict $\ln\gamma_{ij}^\infty$ for any binary mixture of the considered solutes and solvents. In Figure~\ref{Fig:S9} in the Supporting Information, we compare the performance of the whisky method with the data-driven MCM for predicting all available experimental data points. Again, we observe a significant improvement with the proposed whisky method.

In conclusion, we introduce a novel approach to hybridize physical and data-driven prediction methods for physicochemical properties. In this paper, we focused on predicting activity coefficients at infinite dilution, but the approach can directly be transferred to other properties. The proposed method is termed \emph{whisky}, reflecting its similarities with the manufacturing of whisky as it combines model \emph{distillation} with \emph{maturation}. As a Bayesian approach, it incorporates physical knowledge in the form of a prior belief, and allows to combine it with empirical data evidence in a theoretically well-motivated and convenient way. The proposed method outperforms all considered baselines in predicting activity coefficients at infinite dilution in binary mixtures: the physical gold standard UNIFAC Dortmund~\cite{weidlich_1987,constantinescu_2016}, the purely data-driven MCM from our previous work~\cite{jirasek_2020}, and two established machine learning ensemble methods, bagging and boosting. We further show that the whisky method is more robust to outliers in the base methods and has a broader applicability than the hybrid baselines. 
We demonstrate that probabilistic machine learning is perfectly suited for incorporating physical knowledge (that is often abundant in thermodynamics) in powerful data-driven models. We emphasize the generic nature of the proposed whisky approach that opens perspectives to a new generation of hybrid prediction methods for physicochemical properties beyond purely data-driven or purely physical approaches. The transfer to further mixture properties and other physical and data-driven base methods is straightforward. We expect additional improvements if explicit physical information is incorporated and exciting insights by elucidating relations between the learned component features and physical component descriptors.\\

\section*{Acknowledgements}
F.J. greatly acknowledges financial support from the German Academic Exchange Service (DAAD). This material is based upon work supported by the Defense Advanced Research Projects Agency (DARPA) under Contract No. HR001120C0021. Any opinions, findings and conclusions or recommendations expressed in this material are those of the author(s) and do not necessarily reflect the views of the Defense Advanced Research Projects Agency (DARPA). Furthermore, this work was supported by the National Science Foundation under Grants 1928718, 2003237, and by Qualcomm.

\bibliographystyle{unsrt} 
\bibliography{MCM_2} 

\newpage

\section*{Supporting Information}
\renewcommand{\thesection}{S.\arabic{section}}
\renewcommand{\theequation}{S.\arabic{equation}}
\renewcommand{\thefigure}{S.\arabic{figure}}
\renewcommand{\thetable}{S.\arabic{table}}
\setcounter{figure}{0}   
\setcounter{equation}{0}  
\setcounter{table}{0}   
\renewcommand{\theHtable}{Supplement.\thetable}
\renewcommand{\theHfigure}{Supplement.\thefigure}
\renewcommand{\theHequation}{Supplement.\theequation}

\subsection*{Data}
This work is based on the same data set as our previous work\cite{jirasek_2020}; the following overview is therefore rather brief.

Data on activity coefficients at infinite dilution in binary mixtures $\gamma_{ij}^{\infty}$ at $298.15\pm 1$~K were adopted from the Dortmund Data Bank (DDB) 2019~\cite{DDB_2019}. Only molecular components (and one ionic liquid that slipped through our filter) of which the molecular formula is known were considered. Furthermore, metals and data points that are indicated to be of poor quality in the DDB were rejected. If multiple data on $\gamma_{ij}^{\infty}$ for a specific binary mixture $i-j$ in the considered temperature range were available, the arithmetic mean of the available data was used. Furthermore, only components for which at least data in two different binary mixtures are available in the DDB were considered, which is a prerequisite for the application of leave-one-out cross-validation that was applied to evaluate predictive performances of the studied methods. The resulting data set comprises 240 solutes $i$ and 250 solvents $j$. Experimental data on $\gamma_{ij}^{\infty}$ in the considered temperature range are available for 4{,}094 of the 60{,}000 possible binary mixtures. Some components appear as both solute and solvent in the data set. The entries of $\gamma_{ij}^{\infty}$ where $i$ and $j$ denote the same substance, i.e., for pure components, were set to unity to satisfy thermodynamic consistency. These data points were only included in the training data but not considered in the evaluation of the predictive performance. For information on the considered solutes and solvents, we refer to the Supporting Information of our previous work.~\cite{jirasek_2020}

\begin{figure}[htbp]
    \centering
    \includegraphics[width=0.6\textwidth]{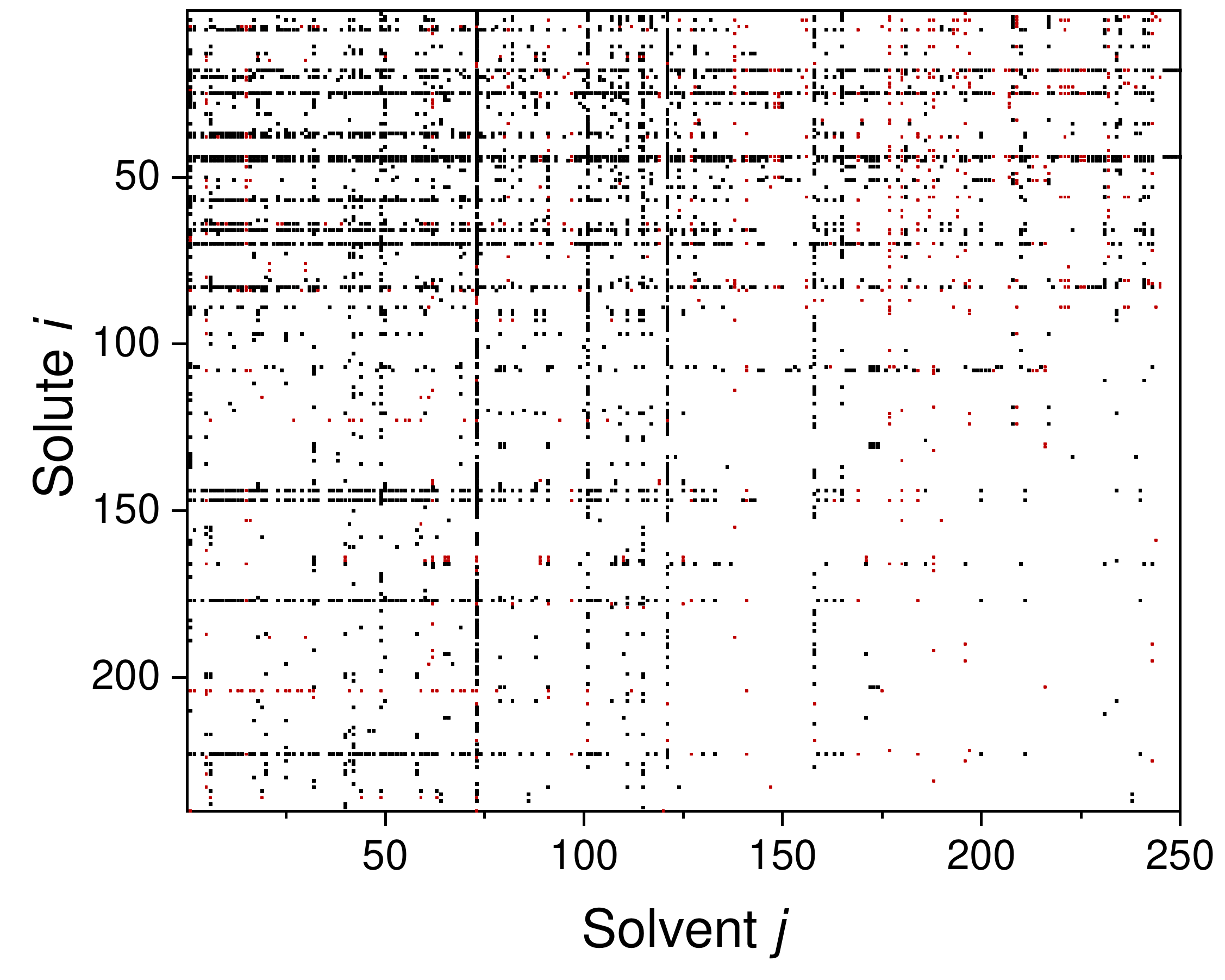}
    \caption{Matrix representation of the available experimental data for $\gamma_{ij}^{\infty}$ at 298.15$\pm$1~K in the DDB~2019~\cite{DDB_2019}. Whitespace: no experimental data available. Black squares: experimental data available, UNIFAC~\cite{constantinescu_2016,weidlich_1987} can be applied. Red squares: experimental data available, UNIFAC cannot be applied.}
    \label{Fig:S1}
\end{figure}

For the comparison of the predictive performances of the different studied methods, 
the data set was further narrowed down since the physical base method, modified UNIFAC (Dortmund)~\cite{weidlich_1987,constantinescu_2016}, referred to as UNIFAC in the following, can (with its present publicly accessible parameterization~\cite{constantinescu_2016}) only be applied to predict $\gamma_{ij}^{\infty}$ for 80\% of the relevant mixtures. Figure~\ref{Fig:S1} shows the matrix representing all possible binary mixtures of the considered solutes and solvents. The color of each entry indicates availability of experimental data in the DDB~\cite{DDB_2019} and applicability of UNIFAC to predict the respective data points (see figure caption).

\clearpage

\subsection*{Model Details}
\subsubsection*{Bayesian Matrix Completion}
As in our previous work~\cite{jirasek_2020}, we used a Bayesian approach to matrix completion to predict $\ln\gamma_{ij}^{\infty}$ (the logarithm of the activity coefficient is used for scaling purposes throughout this work). This approach consists of three steps. In the first step, a generative probabilistic model for the variable of interest, i.e., $\ln\gamma_{ij}^{\infty}$, as a function of initially unknown (latent) features of the solutes $i$ and solvents $j$ is specified. $\ln\gamma_{ij}^{\infty}$ is thereby modeled as the dot product of the feature vector $u_i$ of the solute $i$ and the feature vector $v_j$ of the solvent $j$:
\begin{equation}
    \ln\gamma_{ij}^\infty = u_i \cdot v_j + \varepsilon_{ij}
\end{equation}
where the random variable $\varepsilon_{ij}$ captures both measurement noise and inaccuracies of the model. Both $u_i$ and $v_j$ are vectors of length $K$ containing features of solute $i$ and solvent~$j$, respectively. The hyperparameter $K$ is the number of considered features per component and also called latent dimension, and was set to $K=4$ as in our previous work~\cite{jirasek_2020} for all approaches discussed here.

In the second step, the latent features are inferred by training the generative model to the available data for $\ln\gamma_{ij}^{\infty}$, which requires inverting the generative model. We use Gaussian meanfield variational inference~\cite{blei2017variational,zhang2018advances,kucukelbir2017} for this purpose, which was demonstrated to be robust and efficient in our previous work~\cite{jirasek_2020}. Since our generative model is probabilistic, the inferred latent features are random variables and a probability distribution, called posterior, for each latent feature is obtained. In the third step, we use the means $\mu_{u_i}$ and $\mu_{v_j}$ of the inferred approximate posterior distributions over $u_i$ and $v_j$, respectively, to obtain predictions from the dot product:
\begin{equation}
	\ln(\gamma_{ij}^{\infty})^{\rm pred} = \mu_{u_i} \cdot \mu_{v_j}
\end{equation}
We note that the feature vectors $u$ and $v$ represent a characterization of each solute and solvent, respectively, that is exclusively inferred from the available data for $\ln\gamma^{\infty}$ in binary mixtures of these components. Hence, no explicit physical knowledge on the pure solutes or solvents, e.g., molecular or pure component descriptors like molar mass, dipole moment, or structural formula, was used to find suitable feature vectors. However, since the data for $\ln\gamma^{\infty}$ \textit{implicitly} comprise physical information on the respective components (which is extracted during training the model and aggregated in the feature vectors), relationships between the learned features and physical descriptors of the components can be expected. Preliminary studies have shown that there are no direct correlations between features and physical descriptors. However, to unveil these (complex) relationships will be an exciting task for future work, possibly generating previously unknown physical insights.

All predictions were obtained by leave-one-out cross-validation, i.e., by training the model to all available data except for the one data point that is to be predicted, and repeating this procedure for each available data point. This procedure ensures that the model cannot cheat by training to the test data. In all cases, we used the Stan framework~\cite{carpenter2017}, which allows the specification of user-defined generative models and automates the task of Bayesian inference. The following sections provide implementation details for each of the compared methods. For more information on the theoretical background of Bayesian matrix completion, the reader is referred to our previous work~\cite{jirasek_2020} and the literature~\cite{murphy2012machine}.

\subsubsection*{Data-driven Matrix Completion Method (MCM)}

\begin{figure}[htbp]
    \centering
    \includegraphics[width=\textwidth]{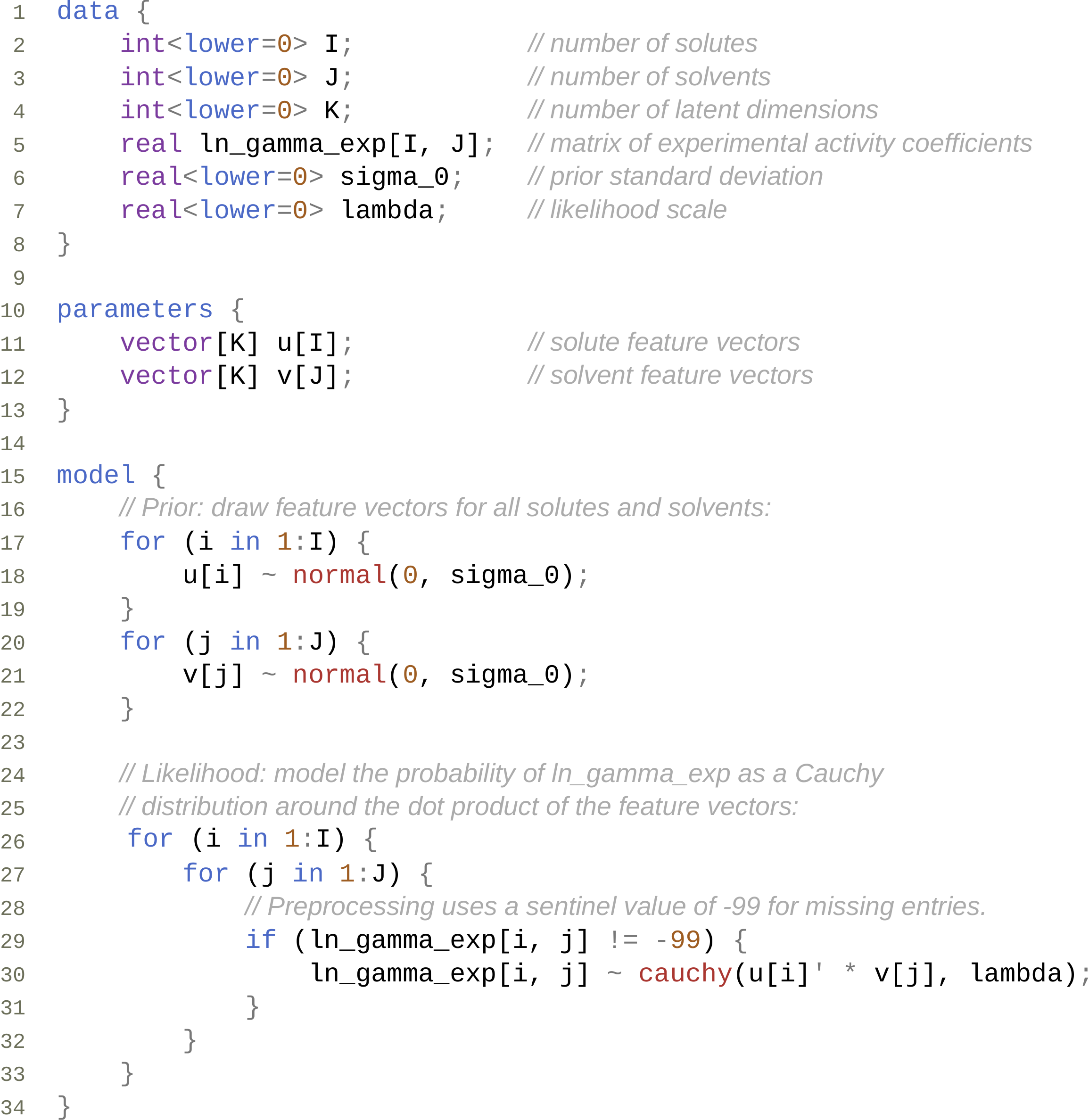}
    \caption{Stan code of the probabilistic generative model of the data-driven MCM. Line 29 ensures that the model is only trained on available experimental data, since missing entries of the matrix were set to -99.}
    \label{Fig:S2}
\end{figure}

Figure~\ref{Fig:S2} shows the Stan code of the probabilistic generative model of the data-driven MCM from our previous work~\cite{jirasek_2020}, which is considered as a data-driven base method here.

For all component features $u_i$ and $v_j$, the same Gaussian prior distribution with mean $\mu_0=0$ and standard deviation $\sigma_0=0.8$ was used. Furthermore, a Cauchy likelihood with scale parameter $\lambda=0.15$ was used for all data points. 
\clearpage

\subsubsection*{Whisky Method}
\label{Model_MCM-pre}
The \emph{whisky} method is proposed in this work as a novel generic approach for hybridizing physical and data-driven prediction methods, and it is applied to predict $\ln\gamma_{ij}^{\infty}$ here. Figure~\ref{flowchart} in the manuscript illustrates the idea of the proposed hybrid approach.

\begin{figure}[htbp]
    \centering
    \includegraphics[width=0.6\textwidth]{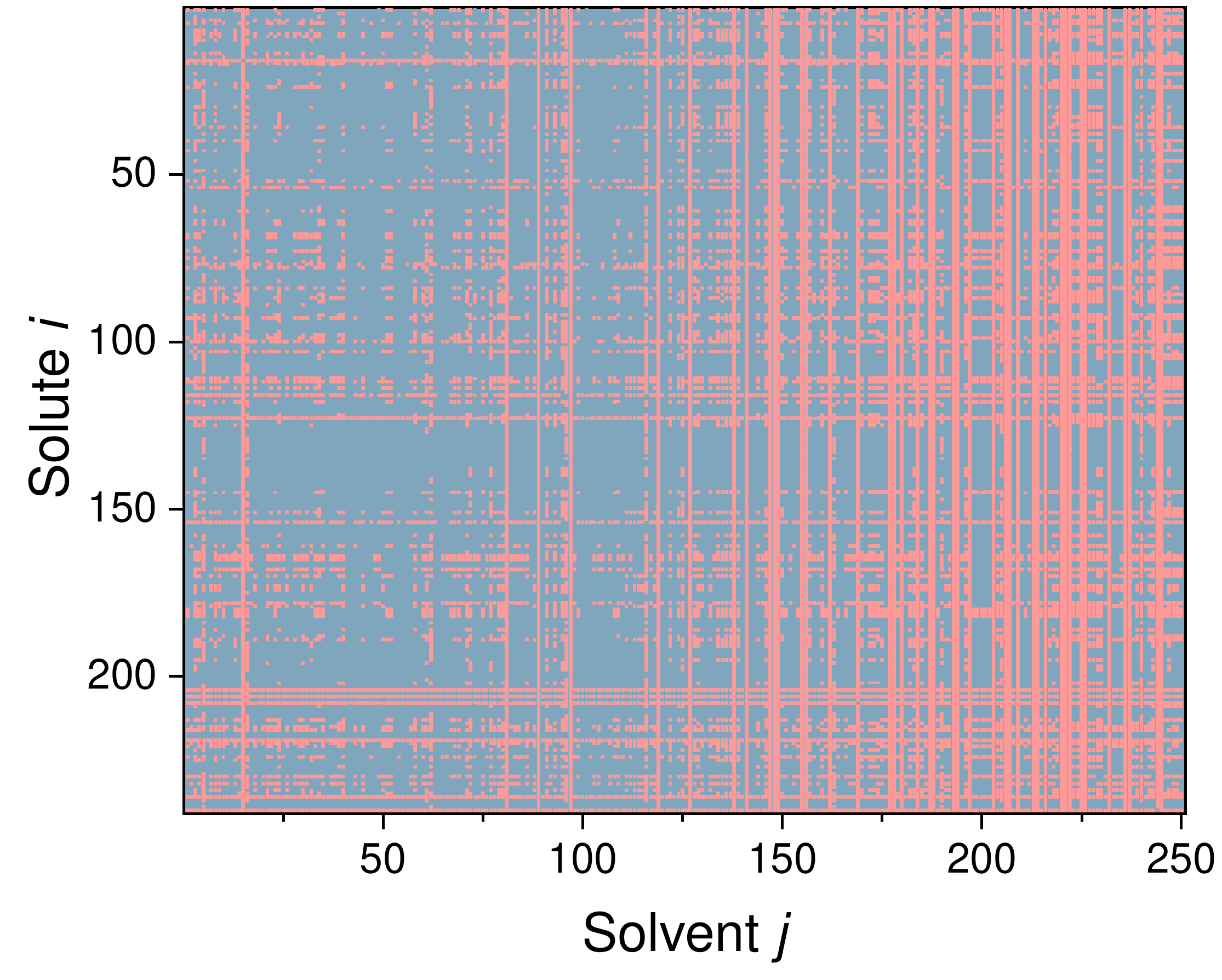}
    \caption{Matrix representation of all possible binary mixtures of the considered solutes $i$ and solvents $j$. Blue: UNIFAC can be applied to predict $\gamma_{ij}^{\infty}$. Red: UNIFAC cannot be applied to predict $\gamma_{ij}^{\infty}$.}
    \label{Fig:S3}
\end{figure}

The method consists of two steps: in the first step (\emph{distillation} step, purple part of Figure~\ref{flowchart} in the manuscript), UNIFAC is employed to predict $\ln\gamma_{ij}^{\infty}$ at 298.15 K in all possible combinations of the considered 240 solutes $i$ and 250 solvents $j$. With the current publicly accessible parameterization of UNIFAC~\cite{constantinescu_2016}, approx. 66\% of all relevant binary mixtures can be modeled. Hence, a rather dense matrix with approx. 66\% observed entries, i.e., UNIFAC predictions for $\ln\gamma_{ij}^{\infty}$, is obtained, cf. Figure~\ref{Fig:S3}.

\begin{figure}[htbp]
    \centering
    \includegraphics[width=\textwidth]{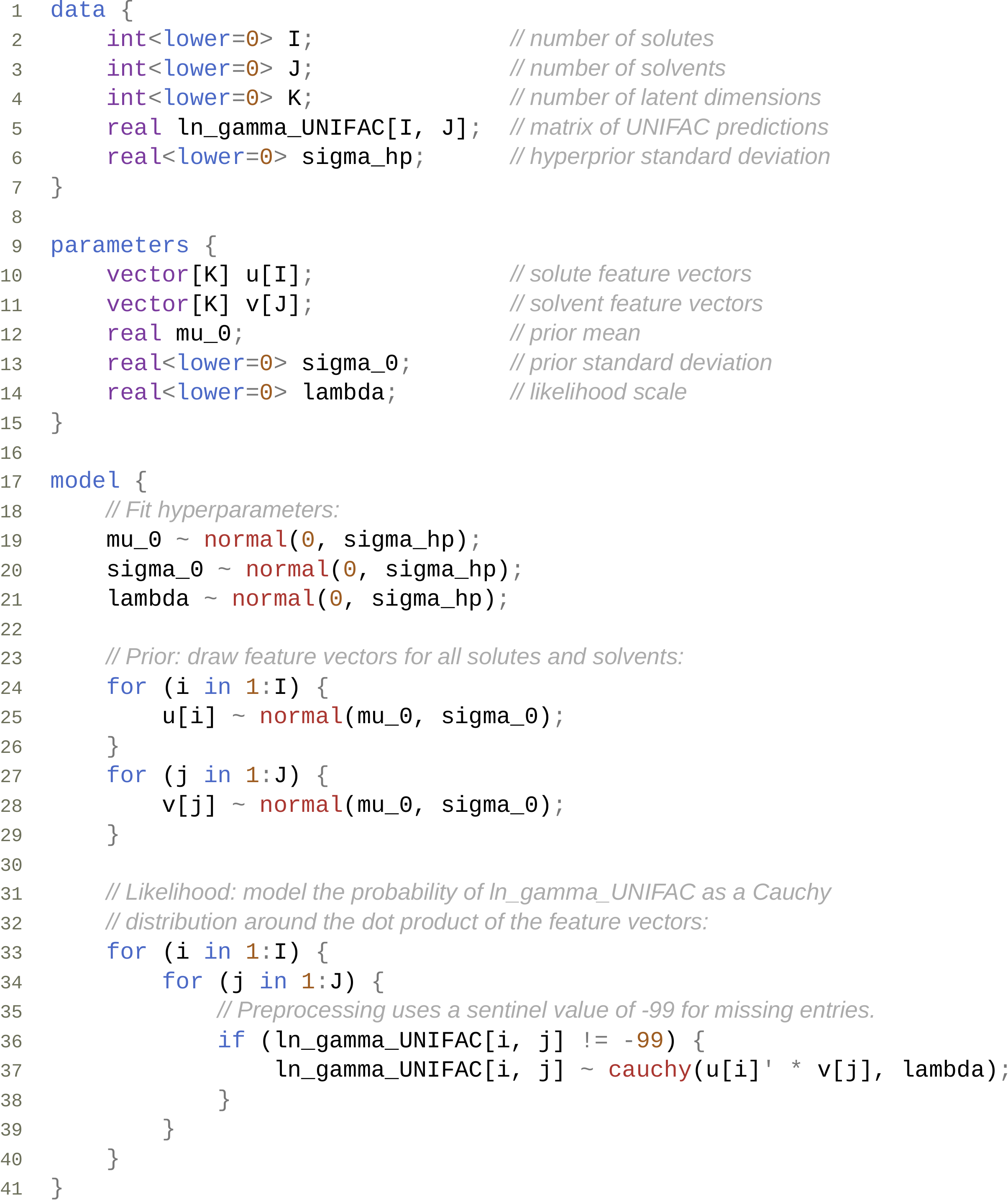}
    \caption{Stan code of the probabilistic generative model of the distillation step of the proposed whisky method. Line 36 ensures that the model in only trained on available UNIFAC predictions, since missing entries of the matrix were set to -99.}
    \label{Fig:S4}
\end{figure}

At first, this rather dense matrix is used for training a Bayesian MCM in the  distillation step. In this step, the parameters (component features) and hyperparameters of the model are trained simultaneously. Therefore, a (strongly uninformative) Gaussian hyperprior with mean $\mu_\mathrm{hp}=0$ and standard deviation $\sigma_\mathrm{hp}=100$ was used for all hyperparameters: the mean $\mu_0$ and standard deviation $\sigma_0$ of the Gaussian prior and the scale parameter $\lambda$ of the Cauchy likelihood. Figure~\ref{Fig:S4} shows the Stan code of the generative model of the distillation step of the whisky method. 

\begin{figure}[htbp]
    \centering
    \includegraphics[width=\textwidth]{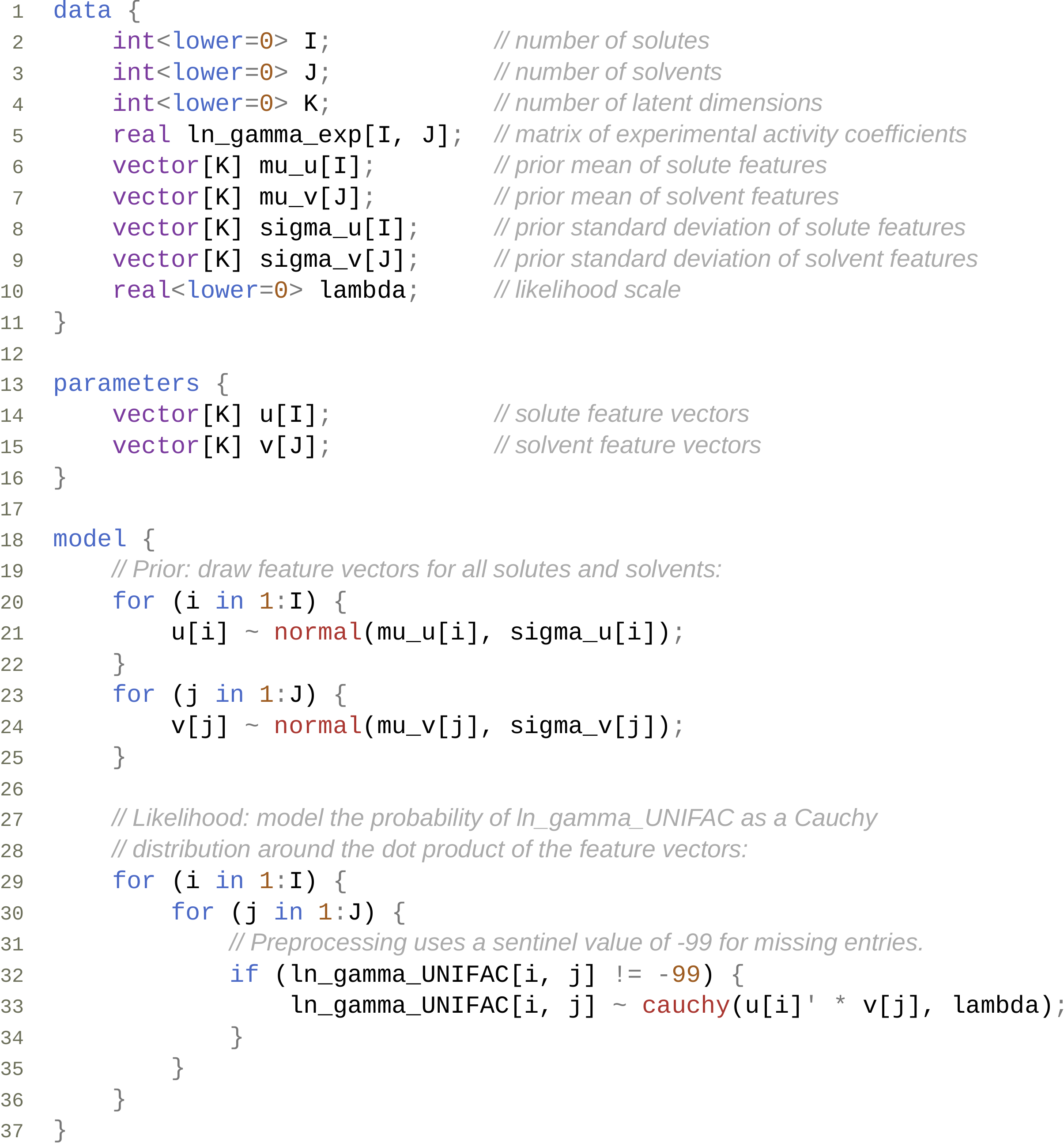}
    \caption{Stan code of the probabilistic generative model of the maturation step of the proposed whisky method. Line 32 ensures that the model is only trained on available experimental data, since missing entries of the matrix were set to -99.}
    \label{Fig:S5}
\end{figure}

The posterior of the distillation step constitutes probability distributions for all component features, thus containing information from the UNIFAC predictions for $\ln\gamma_{ij}^{\infty}$. However, meaningful features were only obtained for the components that can, in principle, be modeled with UNIFAC, i.e., for which UNIFAC predictions were available during the distillation step. These posterior distributions were used to generate informative priors for the subsequent maturation step of the whisky method, the actual training of the approach to the available experimental data (green part of Figure~\ref{flowchart} in the manuscript). In detail, the mean of each posterior distribution of the distillation step was adopted and used in combination with a standard deviation of $\sigma_0=0.5$ in a Gaussian prior in the maturation step. For those components that cannot be modeled with UNIFAC, i.e., for which no UNIFAC predictions were available during the distillation step, a Gaussian prior with mean $\mu_0=0$ and standard deviation $\sigma_0=3$ was used in the maturation step. Hence, we used a rather small prior standard deviation, i.e., a rather strong or informative prior, for those components for which we could extract and hand over information from the distillation to the maturation step. In contrast, we used a rather large prior standard deviation, i.e., a rather weak or uninformative prior, for those components for which no a-priori information could be generated with UNIFAC. The actual numbers for $\sigma_0$ for the informative and uninformative priors in the maturation step are to a certain degree arbitrary but its ratio matches the ratio of the mean and maximum posterior standard deviation of the distillation step. Furthermore, the proposed whisky method is quite robust with respect to $\sigma_0$ in the maturation step. Figure~\ref{Fig:S5} shows the Stan code of the maturation step, the actual training step, of the whisky method.

The proposed whisky method can be applied to predict $\ln\gamma_{ij}^{\infty}$ for any mixture of the considered solutes and solvents. Hence, predictions for the complete data set on $\ln\gamma_{ij}^{\infty}$ from our previous work~\cite{jirasek_2020} that contains 4{,}094 experimental data points and covers 240 solutes $i$ and 250 solvents $j$ can be obtained with the whisky approach. In the manuscript, we only show results for the data points that can also be predicted with UNIFAC for the reason of comparability. However, the performance of the whisky method to predict all available 4{,}094 data points is demonstrated below and compared to the data-driven base method MCM\cite{jirasek_2020}.

\subsubsection*{Bagging}
For obtaining predictions with the bootstrap aggregation (bagging) approach, the arithmetic mean of the predictions of the two base methods UNIFAC~\cite{weidlich_1987,constantinescu_2016} and data-driven MCM~\cite{jirasek_2020} was calculated for each available data point:
\begin{equation}
    \ln(\gamma_{ij}^{\infty})^\mathrm{Bagging}=\frac{1}{2}\left(\ln(\gamma_{ij}^{\infty})^\mathrm{UNIFAC}+\ln(\gamma_{ij}^{\infty})^\mathrm{MCM\mhyphen data}\right)
\end{equation}
For the UNIFAC predictions, an inhouse MATLAB implementation with the latest publicly accessible parameterization~\cite{constantinescu_2016} was used. The MCM predictions were adopted from our previous work~\cite{jirasek_2020}.

With the data-driven MCM, predictions for all 4{,}094 available experimental data points are obtained; with UNIFAC, predictions for only about 80\% of the these can be obtained. Hence, the applicability of the bagging approach is directly limited by the applicability of UNIFAC. The relevant data set covers 231 solutes and 205 solvents.

\subsubsection*{Boosting}
\label{Model_MCM-res}
Additionally, another established machine learning ensemble method, namely boosting, is adopted here as hybrid baseline. The basis for the boosting method is the matrix containing UNIFAC predictions for $\ln\gamma_{ij}^{\infty}$ for the mixtures for which also experimental data on $\ln\gamma_{ij}^{\infty}$ are available. As described above, UNIFAC yields only predictions for about 80\% of the experimental data, which can also be arranged in a partially observed matrix, whose rows and columns correspond to the solutes $i$ and solvents $j$, respectively. Since the results of the boosting method were also evaluated using leave-one-out cross-validation, at least two observed entries, i.e., entries for which experimental data \textit{and} UNIFAC prediction are available, per row and column were required. The data set therefore slightly reduced further, covering 224 solutes and 205 solvents. For all applicable mixtures, the residuals $r_{ij}$ of UNIFAC, i.e., the differences between the experimental data points and the UNIFAC predictions, were calculated:
\begin{equation}
    r_{ij}=\ln(\gamma_{ij}^{\infty})^{\mathrm{exp}}-\ln(\gamma_{ij}^{\infty})^{\mathrm{UNIFAC}}
\end{equation}

\begin{figure}[htbp]
    \centering
    \includegraphics[width=\textwidth]{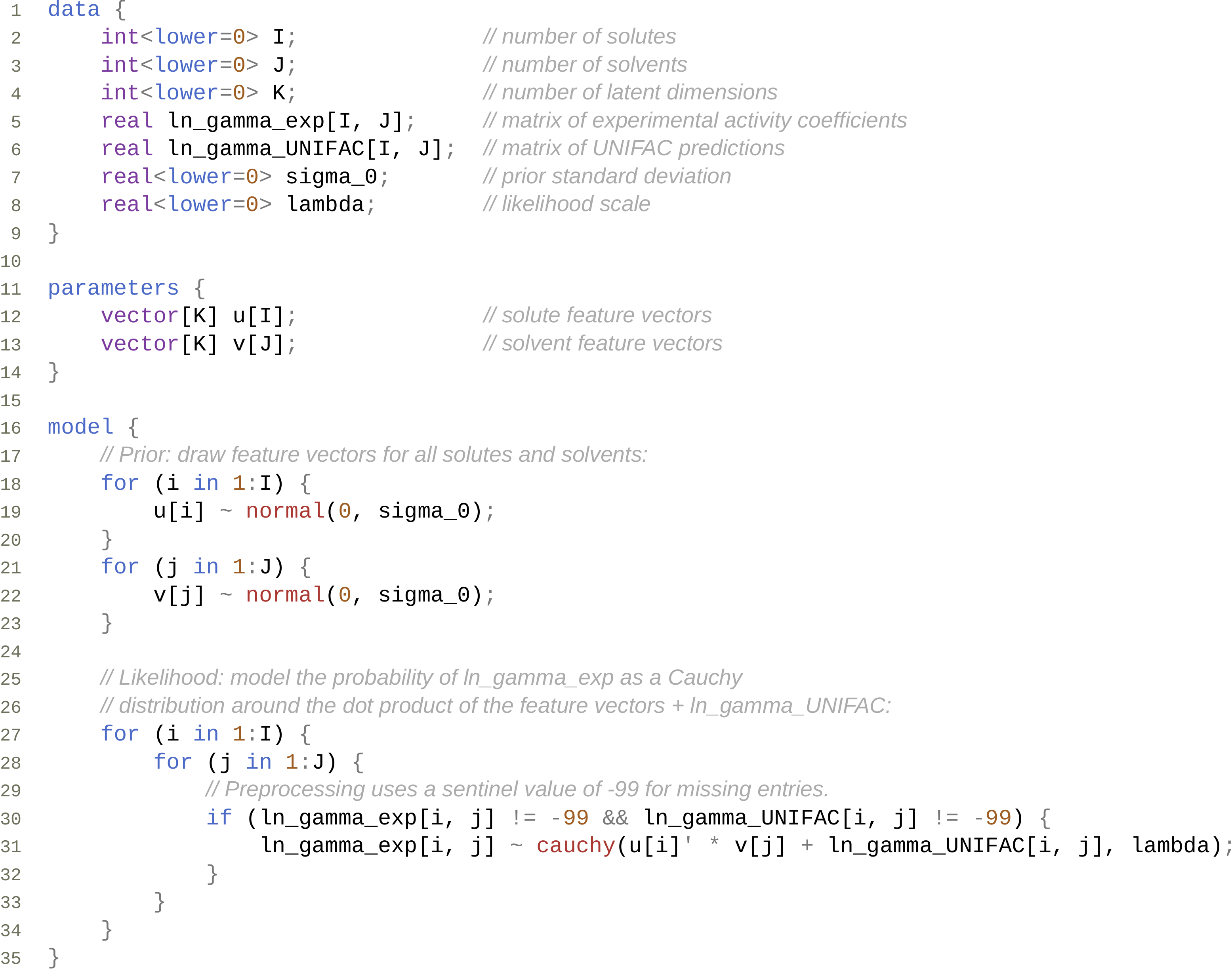}
    \caption{Stan code of the probabilistic generative model of the boosting method. Line 30 ensures that the model is only trained on available data, since missing entries of the matrix were set to -99.}
    \label{Fig:S6}
\end{figure}

These UNIFAC residuals were arranged in a partially observed matrix to which the concept of Bayesian matrix completion was applied in the second step. Hence, previously unknown features of the solutes and solvents that describe the deviation of the UNIFAC predictions from the experimental data were learned from the training data. For all parameters a Gaussian prior with mean $\mu_0=0$ and standard deviation $\sigma_0=1$ was used here. Furthermore, a Cauchy likelihood with scale parameter $\lambda=0.15$ was used. Figure~\ref{Fig:S6} shows the Stan code for the boosting method.

Following the concept of leave-one-out cross-validation, the solute and solvent features were trained to all $r_{ij}$ except for one, which was then considered as test data point and predicted. Each data point served once as test data point. The prediction of the logarithmic activity coefficient $\ln(\gamma_{ij}^{\infty})^\mathrm{Boosting}$ was calculated in a straightforward manner:
\begin{equation}
    \ln(\gamma_{ij}^{\infty})^\mathrm{Boosting}=\ln(\gamma_{ij}^{\infty})^\mathrm{UNIFAC} + \mu_{u_i} \cdot \mu_{v_j}
\end{equation}
where $\ln(\gamma_{ij}^{\infty})^\mathrm{UNIFAC}$ is the UNIFAC prediction for the activity coefficient, and $\mu_{u_i}$ and $\mu_{v_j}$ are the posterior means of the corresponding solute and solvent feature vectors, respectively, obtained from the MCM of the residuals $r_{ij}$.

\clearpage

\subsection*{Additional Results}
\subsubsection*{Results of Bagging and Boosting}

\begin{figure}[h]
    \centering
    \includegraphics[width=0.45\textwidth]{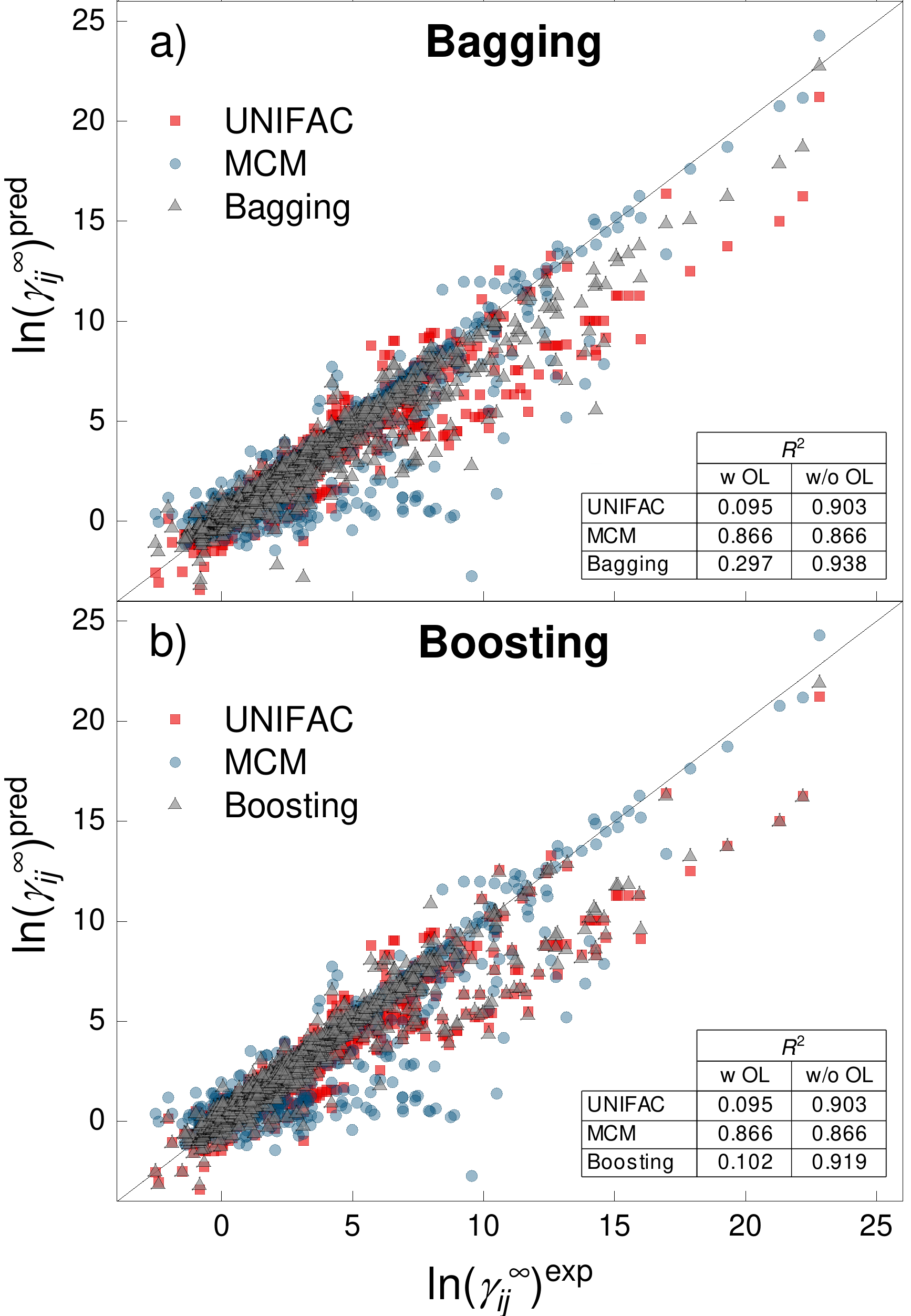}
    \caption{Parity plots of the predictions (pred) for $\ln\gamma_{ij}^\infty$ with the bagging (a) and boosting (b) approaches over the corresponding experimental values (exp) and comparison to UNIFAC and data-driven MCM. Coefficients of determination~$R^2$ (higher is better, $1$ implies perfect correlation) are given, both including and excluding the worst eight UNIFAC outliers (OL).}
    \label{Fig:S7}
\end{figure}

In Figure~\ref{Fig:S7}, the predictions with the hybrid baselines bagging and boosting are represented in parity plots and compared to the predictions of the base methods UNIFAC and data-driven MCM. Both hybrid approaches only partially compensate for outliers of the data-driven MCM but severely suffer from outliers of UNIFAC. The worst outliers of UNIFAC, bagging, and boosting lie outside the depicted ranges, cf. Figure~\ref{Fig:S8}. Slight improvements with bagging and boosting compared to the data-driven MCM can only be achieved, if the worst UNIFAC outliers are ignored, cf. the coefficients of determination $R^2$ in Figure~\ref{Fig:S7} and Figure~\ref{Comparison} in the manuscript. 

The slight improvements that are possible with the bagging approach compared to MCM if (and only if) the worst UNIFAC outliers are ignored can mainly be attributed to error cancellations by averaging the predictions of the two base methods (UNIFAC and MCM). For approx. 67\% of the applicable data points, better predictions are obtained with bagging than with UNIFAC, whereas for approx. 52\% of the data points, the predictions with bagging are better than those of the data-driven MCM. Hence, the bagging approach improves the predictions of both base methods for more data points than impairs the predictions. This can be explained by the observation that for almost 19\% of the data points, an improved prediction accuracy compared to \textit{both} UNIFAC and MCM is observed. Hence, for these 19\%, the bagging approach benefits from the effect of error cancellation, as the respective data points are overestimated by MCM and underestimated by UNIFAC, or vice versa. Incidentally, it is in the nature of the bagging approach that it cannot impair the predictions of UNIFAC \textit{and} MCM for a specific data point at the same time.

However, in all cases, i.e., with or without ignoring the worst UNIFAC outliers, the performances of bagging and boosting are significantly worse than the performance of the proposed whisky method, cf. Figure~\ref{Comparison} and \ref{Parity_Plot} in the manuscript. 

\clearpage

\subsubsection*{UNIFAC Outliers}

\begin{figure}[htb]
    \centering
    \includegraphics[width=0.45\textwidth]{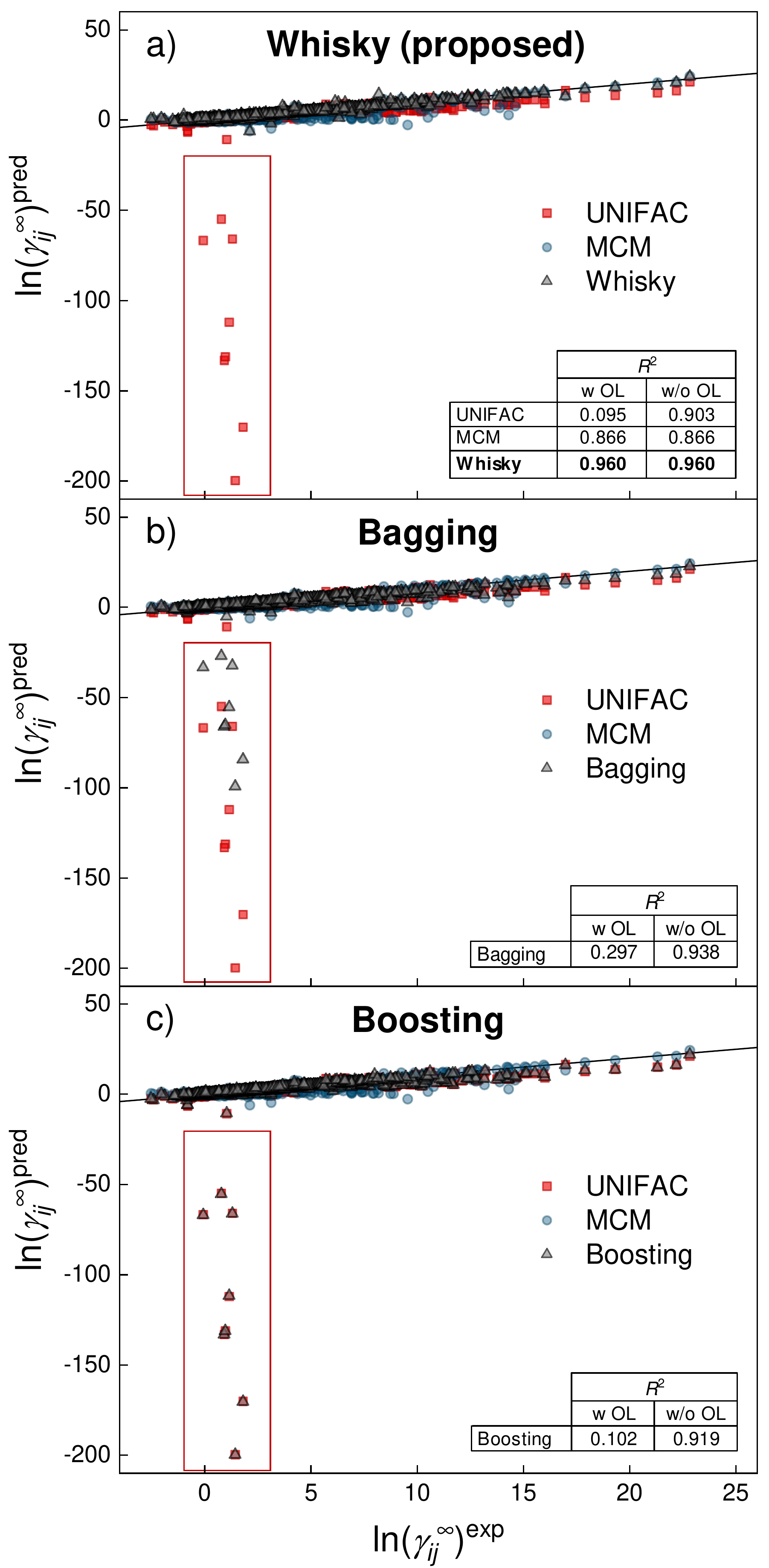}
    \caption{Parity plots of the predictions (pred) for $\ln\gamma_{ij}^\infty$ with the hybrid approaches over the corresponding experimental values (exp) and comparison to UNIFAC and MCM. a) whisky (proposed), b) bagging, c) boosting. The worst eight UNIFAC outliers (OL) are marked by red boxes. Coefficients of determination $R^2$ (higher is better, $1$ implies perfect correlation) are given, both including and excluding OL.}
    \label{Fig:S8}
\end{figure}

In Figure~\ref{Fig:S8}, predictions for all applicable $\ln\gamma_{ij}^\infty$ \textit{including} the worst eight UNIFAC outliers (OL) with all studied methods are shown in a parity plot representation. For these UNIFAC outliers, marked by red boxes in Figure~\ref{Fig:S8}, the hybrid methods bagging and boosting, shown in panels b) and c), respectively, give poor predictions, while the whisky method, shown in panel a), as well as the purely data-driven MCM are demonstrated to be much more robust and do not exhibit such outliers.

The performance of UNIFAC strongly depends on the quality of the fitted binary group-interaction parameters. The worst eight outliers of UNIFAC are associated to binary mixtures of a cyclic alkane, cyclic alkene, or furane as solute and a heterocyclic compound in which the ring structure is formed by carbon and nitrogen as solvent. According to the group-division scheme of UNIFAC~\cite{weidlich_1987,constantinescu_2016}, all aforementioned solutes contain at least one `CY-CH2' main group (UNIFAC main group no. 42), all aforementioned solvents contain at least one `PYRIDINE' main group (UNIFAC main group no. 18). We therefore attribute the poor predictions of UNIFAC for these outliers mainly to the UNIFAC group-interaction parameters between the `CY-CH2' and the `PYRIDINE' group, which have presumably been overfitted to parts of the data that were used for training UNIFAC (and that are not necessarily part of the data set considered here). We assume that this is not an isolated case but likely to occur for other group-interaction parameters for UNIFAC as well, depending on the data sets that are studied. As the data-driven MCM is an orthogonal approach to UNIFAC, it is conclusive that it does not suffer from the same difficulties. Furthermore, the prior distributions in the probabilistic approach of the data-driven MCM serve as regularization terms and can therefore be expected to prevent overfitting of the MCM to single data points or parts of the data set.

The proposed hybrid whisky approach, like the data-driven MCM, does not exhibit such outliers, although it, in contrast to the data-driven MCM, considers information from UNIFAC. In whisky, information from UNIFAC is taken into account in the prior distributions used in the \textit{maturation} step, cf. Figure~\ref{flowchart} in the manuscript. The nonzero variance of the Gaussian priors that we have used here allows the whisky method to `correct' poor UNIFAC predictions by combining them with experimental data evidence in the maturation step. This procedure seems to work extremely efficiently. Ultimately, we emphasize again that even with omitting these systematic outliers of UNIFAC in the evaluation, the proposed whisky method significantly outperforms the individual and hybrid baselines studied here.

\subsubsection*{MCM and Whisky Predictions for All Available Data}
\begin{figure}[!htb]
    \centering
    \begin{minipage}[t]{0.47\textwidth}
    \vspace{0pt}
        \centering
        \includegraphics[width=\textwidth]{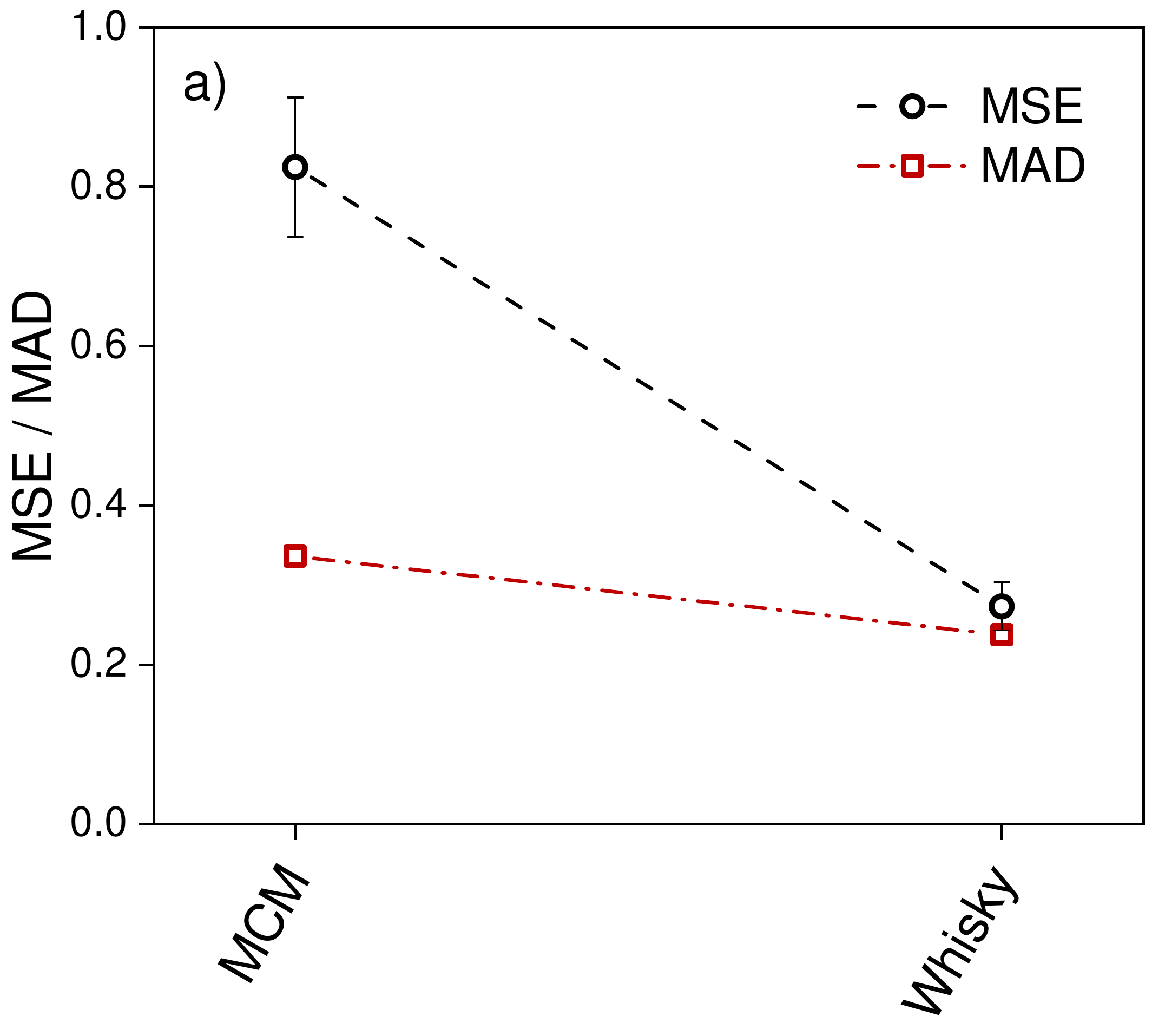}
    \end{minipage}%
    \hfill
    \begin{minipage}[t]{0.48\textwidth}
        \centering
        \vspace{0pt}
        \includegraphics[width=\textwidth]{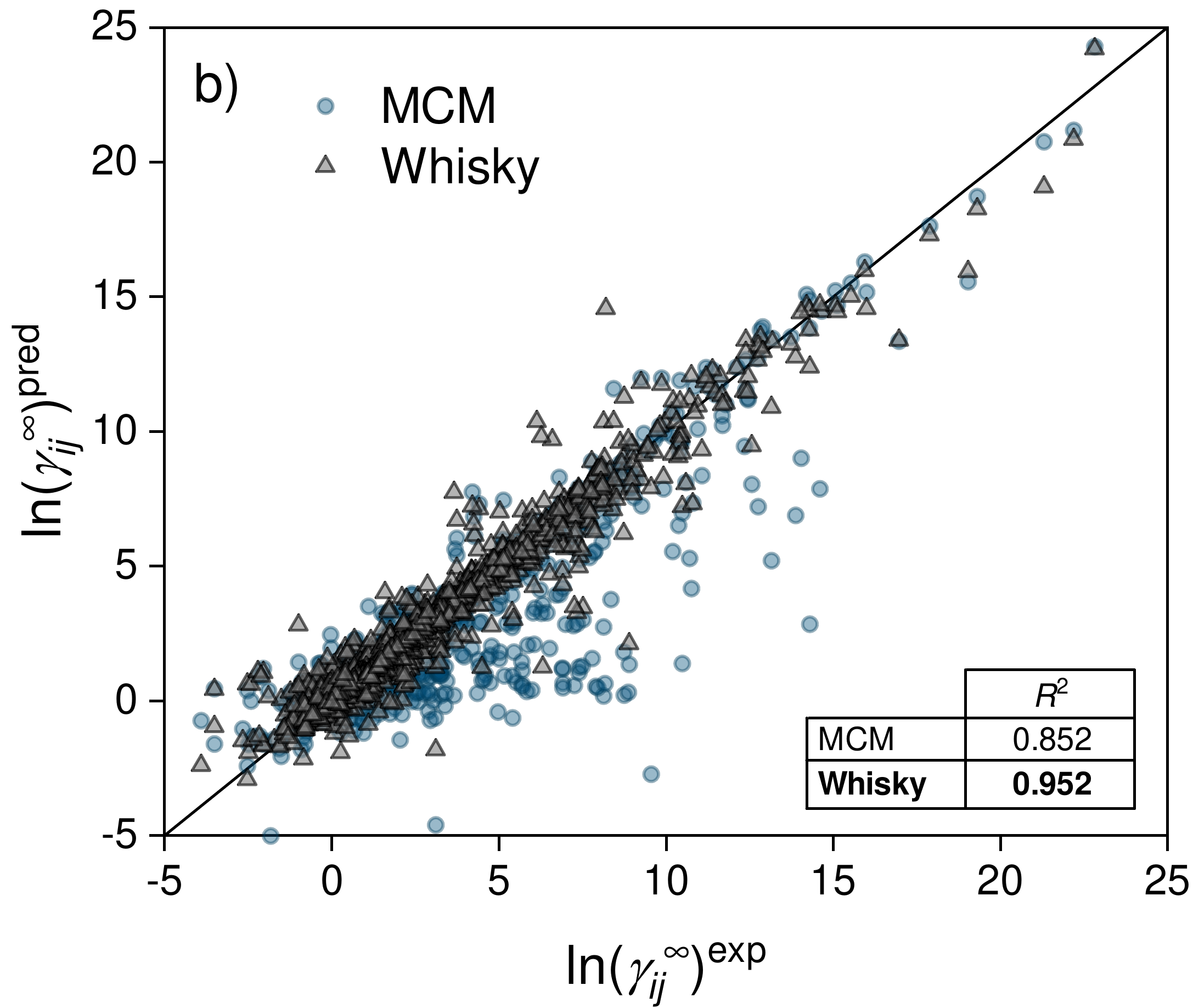}
    \end{minipage}
\caption{Comparison of the data-driven MCM and the proposed whisky method considering all 4{,}094 available experimental data points for $\ln\gamma_{ij}^\infty$. a) Mean square error (MSE) and mean absolute deviation (MAD) of the predictions; lower is better for both metrics, error bars represent the standard errors of the means. b) Parity plot of the predictions (pred) over corresponding experimental values (exp) and coefficients of determination $R^2$ (higher is better, $1$ implies perfect correlation).}
\label{Fig:S9}
\end{figure}

As described above, the whisky method yields predictions for all  available experimental $\ln\gamma_{ij}^\infty$ for the considered solutes $i$ and solvents $j$, while the bagging and boosting methods are limited by the applicability of UNIFAC. In Figure~\ref{Fig:S9}, the performance of the whisky approach to predict $\ln\gamma_{ij}^\infty$ for all 4{,}094 available data points is demonstrated and compared to the performance of the data-driven MCM. Significant improvements with respect to mean square error (MSE), mean absolute deviation (MAD), and coefficient of determination ($R^2$) are obtained with the whisky method. 

\subsubsection*{Influence of Latent Dimension}

\begin{figure}[htb]
    \centering
    \includegraphics[width=0.5\textwidth]{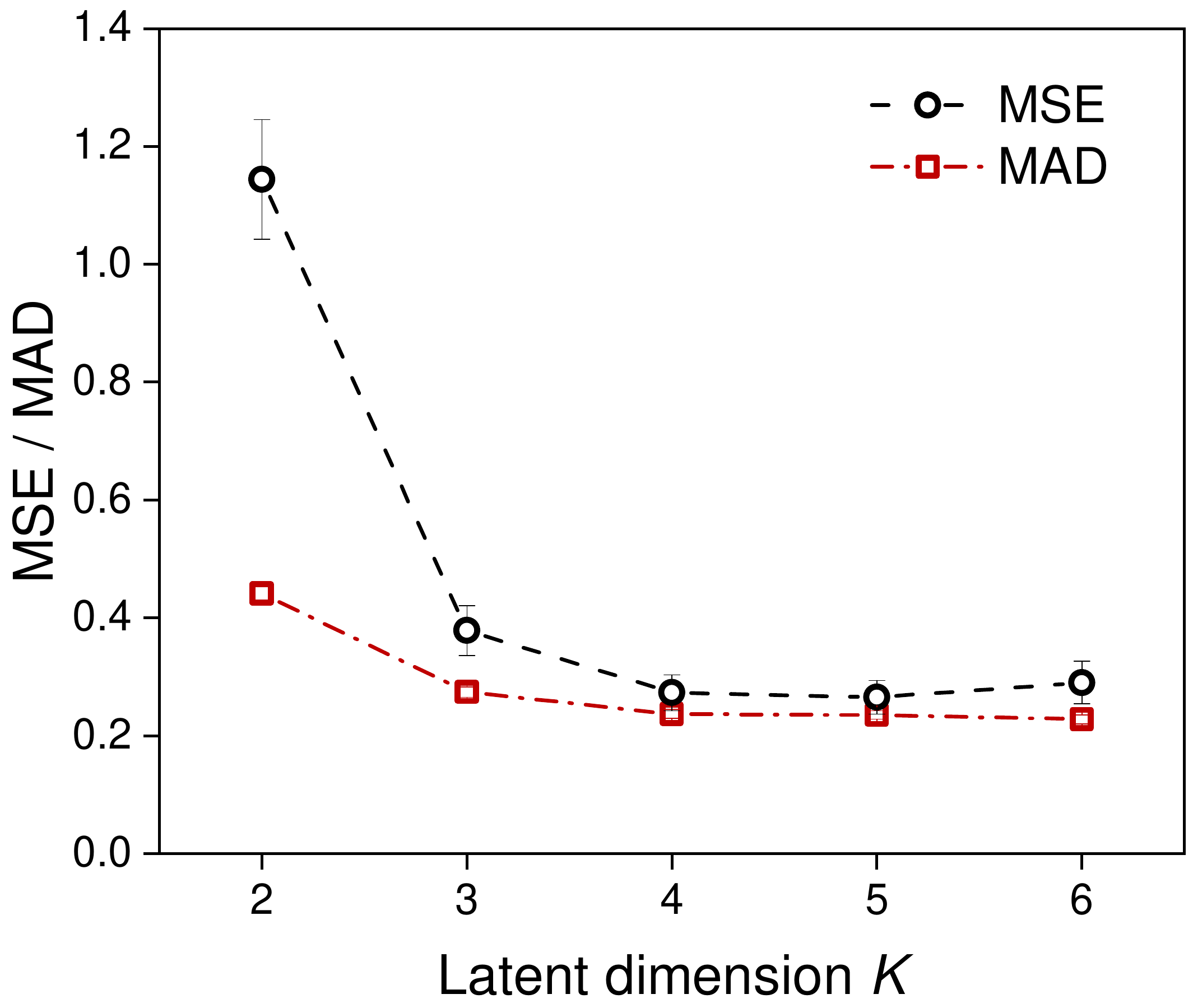}
    \caption{Influence of the latent dimension $K$ on mean square error (MSE) and mean absolute deviation (MAD) of the predictions with the proposed whisky method for all 4{,}094 available experimental data points for $\ln\gamma_{ij}^\infty$; lower is better for both metrics, error bars represent the standard errors of the means.}
    \label{Fig:S10}
\end{figure}

Figure~\ref{Fig:S10} shows MSE and MAD scores that are obtained with the whisky method considering all available experimental data points for different latent dimensions $K$, specifically for varying numbers of learned solute and solvent features ranging from two to six. For $K=4$, which was used to obtain all other results throughout this work (cf. Section `Model Details' in the ESI), and $K=5$, very similar scores are obtained. Also the scores for $K=6$ are similar, which indicates that the whisky method is rather robust to small enlargements of $K$. However, the slightly worse MSE score for $K=6$ indicates commencing overfitting at larger numbers for $K$. Hence, at very large numbers of $K$, the method is likely to overfit to the training data dropping predictive performance (on unseen test data) due to too high flexibility. On the other hand, also lowering the number of $K$ can result in significant deterioration of the scores. For $K=3$, only slightly worse scores, for $K=2$, substantially worse scores are observed. This indicates that for very small numbers of $K$, the approach is insufficiently flexible for describing the data well. Hence, only two features per solute and solvent are not adequate for characterizing the components well, which results in poor scores due to underfitting.

The optimal number of $K$ strongly depends on the data that are considered and can, if necessary, be determined by cross-validation. In this work, we have simply adopted $K=4$ from our previous work in which we introduced the data-driven MCM and which is also a good choice for the whisky method as demonstrated in Figure~\ref{Fig:S10}. We consider the observed robustness towards small variations of $K$ ($K=3$ to $K=6$ here) as a major strength of the whisky method, as it shows excellent predictive performance on $\ln\gamma^\infty$ data without requiring extensive hyperparameter optimization.

We note that for applying the whisky method to other physicochemical properties than activity coefficients, significantly different numbers for the latent dimension $K$ might be required. I.e., the number of features that are required for adequately characterizing components with regard to a specific property is likely to depend on the considered property. The whisky method is not restricted to specific numbers of $K$; in principle any number can be chosen, which can be determined by cross-validation to prevent both under- and overfitting as described above. Larger numbers of $K$ might hamper the computation time required to train the method, which is, however, for $K=4$ and the data set considered here, in the range of seconds or minutes (using a custom laptop).

\end{document}